\documentclass[10pt,twocolumn,twoside,submit]{JCNtran}
\usepackage{color}
\usepackage{amsmath}
\usepackage{amssymb}
\usepackage{tabularx}
\usepackage{epsfig}
\usepackage{graphicx}
\usepackage[T1]{fontenc}
\usepackage{epstopdf}
\usepackage{subfigure}

\def\BibTeX{{\rm B\kern-.05em{\sc i\kern-.025em b}\kern-.08em
    T\kern-.1667em\lower.7ex\hbox{E}\kern-.125emX}}

\begin{document}
\bibliographystyle{jcn}

\title{Impact of Node Speed on Throughput of Energy-Constrained Mobile Networks with Wireless Power Transfer}
\author{Seung-Woo Ko and Seong-Lyun Kim
\thanks{S.-W.~Ko is with the Department of Electrical and Electronic Engineering, The University of Hong Kong, Pok Fu Lam, Hong Kong (e-mail: swko@eee.hku.hk).
S.-L.~Kim is with the School of Electrical and Electronic Engineering,
Yonsei University, Seoul, Korea
(email: slkim@ramo.yonsei.ac.kr).}
\thanks{S.-W.~Ko is  the corresponding author.}
}
\maketitle

\begin{abstract}
A wireless charging station (WCS) transfers energy wirelessly to nodes within its charging range.
This paper investigates the impact of node speed on throughput of WCS overlaid mobile ad-hoc networks (MANET)
when packet transmissions are constrained by energy status of each node.
Nodes in such a network shows twofold charging pattern depending on their moving speeds.
A slow moving node outside WCS charging regions resorts to wait energy charging from WCSs for a long time
while that inside WCS charging regions consistently recharges the battery.
A fast moving node waits and recharges directly contrary to the slow moving node.
Reflecting these phenomena, we design a two-dimensional Markov chain
of which the state dimensions respectively represent remaining energy and distance to the nearest WCS normalized by node speed.
Solving this enables to provide the following three impacts of speed on throughput.
Firstly, higher node speed improves throughput by reducing the inter-meeting time between nodes and WCSs.
Secondly, such throughput improvement by higher speed is replaceable with larger battery capacity.
Finally, we prove that the throughput scaling is independent of node speed.
\end{abstract}


\vspace{10pt}
\section{\uppercase{Introduction}}
\label{sec:introd}

\subsection{Motivation}
Wireless mobile devices are currently pervasive, and the number of the devices is expected to be ever-growing
when internet-of-things (IoT) and wearable devices emerge in the near future.
This tendency makes their energy supply not only huge but also frequent 
that the existing wired charging technologies cannot cope with.
Faced with the energy supply problem, wireless power transfer (WPT) is fast becoming recognized as a viable solution \cite{Lin2013}.
A node can recharge its battery without plugs and wires if there is an apparatus enabling to perform WPT, known as a {\it wireless charging station} (WCS).

\begin{figure} [t!]
\centering
\includegraphics[angle=0,width=3.00in]{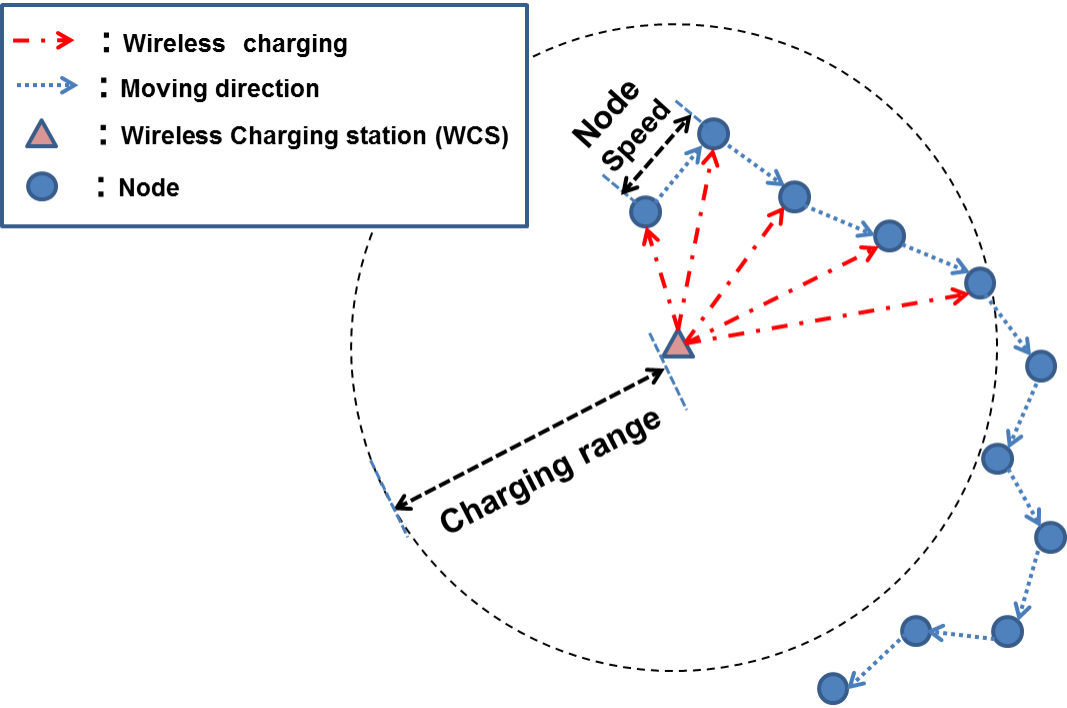}
\caption{The pattern of wireless charging when node speed is slow.
During the period that a node is in the charging region of the WCS,
it receives energy from WCS continuously. Once a node is out of the charging range,
on the other hand, it takes a long time to receive energy from WCS again.
} \label{Fig1}
\end{figure}

This paper deals with the throughput of wireless networks
when WCSs are deployed.
Mobile devices are recharged by WCS via {\it magnetic resonance coupling} \cite{Kurs2007}
of which the efficiency is high only within a few meters.
A node receives energy only when it is located in the said charging region of WCS. 
The throughput of the IoT device is thus greatly influenced by its energy status 
that depends on its mobility pattern, especially the moving speed,   
determining how frequently it can visit WCSs.
Fig.~\ref{Fig1} shows a graphical illustration explaining the impact of speed.
When a node moves slowly, it remains in the charging region of the WCS and
can receive energy from the WCS sequentially.
Once it is out of the region, on the other hand, it takes a long time to receive energy again.
In other words, an irregular energy provision occurs such that some devices consistently receive energy from WCSs while others suffer from the lack of energy supply, encouraging to revisit the throughput of wireless powered mobile networks. 

\subsection{Prior Works}
The most common WPT method is the magnetic inductive coupling
that electric power is delivered by means of an induced magnetic field.
The drawback of this method is its power transfer efficiency that diminishes significantly
unless the transmitter and the receiver are close in contact.
Recently, there have been efforts to develop WPT technology 
of which the efficiency remains high within several-meters range.
In \cite{Kurs2007}, Kurs {\it et al.} suggested a novel method called magnetic resonant coupling 
where electric power is transferred from one to the other with high efficiency when two devices are tuned to the same resonant frequency.
However, its high efficiency requirement is so tight that 
it is vulnerable to the misalignment between a transmitter and a receiver.
Some sophisticated tracking and alignment techniques are proposed for practical use, 
i.e. frequency matching \cite{Sample2011}, impedance matching \cite{Beh2013}
and resonant isolation \cite{Yoon2012}. 
By exploiting them, magnetic resonant coupling is well adapted to mobile environments, and
enables to recharge not only small electronic devices but also large electric vehicles \cite{Witricity}.


There are several studies incorporating WPT into MANET.
In \cite{Xie2012_1} and \cite{Xie2012_2}, the authors suggested a wireless charging vehicle (WCV) that
visits all nodes to recharge their battery, and found the optimal travel path to avoid the battery depletion of each node.
Li {\it et al.} introduced a Qi-ferry \cite{Li2012}, which is similar to WCV except the fact Qi-ferry consumes its own residual energy when it is moving.
In other words, longer travel distance of Qi-ferry visits more nodes whereas accelerates its energy depletion.
They optimized its travel path reflecting the above tradeoff.
{A distributed WPT scheme is proposed in \cite{Madhja2015} where multiple mobile chargers wirelessly provide energy to sensors 
by exploiting the limited network information.}
These papers \cite{Xie2012_1}--\cite{Madhja2015} are based on the assumption that
WPT-enabled devices have knowledge of full or limited geographical information for all rechargeable nodes,
hardly estimated in mobile environments.

In \cite{Huang2013}, Huang analyzed the performance of an energy-constrained mobile network assuming
the energy arrival process of each node as an independent and identically distributed (i.i.d.) sequence, 
which is relevant when many WCSs are employed and the moving speed of each node is fast.    
In \cite{He2013}, He {\it et al.} derived a necessary condition of the number of WCSs needed 
to continue the operation of each node. 
In \cite{Dai2015}, Dai {\it et al.} derived Quality of Energy Provisioning (QoEP), 
the expected portion of time a node sustains its operation.
They show that QoEP converges to one as battery capacity or node speed increases.  
Their analysis is based on the spatial distribution of nodes. 
Since various mobility models follow the same spatial distribution, 
only the lower and upper bounds of QoEP are provided.
{A Markovian mobility model is utilized in \cite{Niyato2015} and \cite{Niyato2014} where a node can move to a few finite points 
according to predetermined transition probabilities,
enabling to study delay-limited and delay-tolerant communications, respectively. 
An intentional movement to a location providing WPT caused by the motivation of battery charging, 
called a spatial attraction, is studied in \cite{Kim2016} showing that the coverage rate can be improved by the optimally controlled power and charging range. }

The aforementioned prior works overlook the impact of node speed affecting the process of energy arrival significantly, thereby making an impact on throughput of the energy-constrained network 
where a packet transmission is constrained by the energy status of each node.
This paper aims at establishing the relationship between the node speed and the throughput in a mathematical manner. 
To the best of our knowledge,  there is no work on figuring out the above impact.


 \subsection{Contributions and Organization}
To investigate the impact of node speed on energy provision and corresponding throughput, 
we develop a new framework using a two-dimensional Markov chain.
Its horizontal and vertical state dimensions respectively represent the remaining energy
and the distance to the nearest WCS.
We derive its steady-state probabilities and express throughput as a function of node speed.
The main contributions of this paper are summarized below.
\begin{itemize}
\item Higher node speed reduces the frequency of lengthy inter-meeting times between a node and a WCS 
and eventually  improves the throughput.
The inter-meeting time is interpreted as an energy-starving duration. 
We explain the phenomenon through the stochastic distribution of the inter-meeting time in Proposition~1.
 
\item 
A slow-moving node stays in the charging region for a long time.
It saves enough energy to endure a lengthy inter-meeting time if its battery capacity, 
the maximum amount of energy stored in the battery, is large enough.   
In Proposition~2, we show that a slow-moving node achieves the same throughput performance as a fast moving one
when the battery capacity becomes infinite.

\item 
In Proposition~3, we show that the throughput scaling is calculated as
$\Theta\left(\min\left(1, \frac{m}{n}\right)c^{\min\left(1,\frac{m}{n}\right)}\right)
$\footnote{We recall that the following notation:
(i) $f(n)=O(g(n))$ means that there exists a constant $c$ and integer $N$ such that $f(n) \leq c g(n)$ for $n >N$.
(ii) $f(n)=\Theta(f(n))$ means that $f(n)=O(g(n))$ and $g(n) = O(f(n))$.}
where $n$ and $m$ respectively denote the number of nodes and WCSs, and $c$ is a constant ($0<c<1$).
As the network becomes denser,
the throughput solely depends on the ratio $\frac{m}{n}$ and becomes independent of node speed unless nodes are stationary.
\end{itemize}

Note that the approach in this work is similar to that of our previous work \cite{Ko2013} as both apply a Markov chain 
to model an energy-constrained mobile network.
In \cite{Ko2013}, it is assumed that nodes follow the i.i.d. mobility model, 
allows us to include only the residual energy status as a Markov chain state.  
On the other hand, our current work focuses on finite node speed, which limits node movement 
within a restricted area.   
In other words, the current node location depends on the previous one. 
Therefore, we should express not only the residual energy, but also the location information of a node when designing a Markov chain model. 
Our paper illustrates that the throughput under the i.i.d. mobility model in \cite{Ko2013} can be understood as an upper bound of that under the finite node speed.
This upper bound is achievable when i) node speed becomes faster, ii) battery capacity becomes larger or iii) node density increases.


The rest of this paper is organized as follows:
In Section II, we explain our models and metrics.
In Section III, we introduce a two-dimensional Markov chain design and derive its steady state probabilities.
In Section IV, we verify how the impact of node speed on the throughput is influenced by battery capacity and node density.
Finally, we conclude this paper in Section V.

\vspace{10pt}
\section{\uppercase{Models and Metrics}}\label{sec:model}

\subsection{Network Description}

{Consider a wireless network where $n$ nodes and $m$ WCSs are randomly located in a torus area of size $\sqrt{S}\times \sqrt{S}$ square meter.}
Time is divided into equal length slots.  
In each  slot, a node randomly changes its direction and moves at a speed of $v$ (meter$/$time slot).
Therefore, we have:
\begin{eqnarray}\label{inequality}
\| X_i(t+1)-X_i(t)\| = v,
\end{eqnarray}
where $X_i(t)$ is the location of node $i$ at slot $t$ and $\|\cdot\|$ means the Euclidian distance. 
The purpose of this mobility modelling is to focus on the impact of node speed, the primary issue of this paper.
Although this model may not be entirely realistic,
it enables us to develop a tractable approach placing emphasis on node speed.

A node can transmit its packet to one of neighbors within transmission range $r$.
According to the {\it protocol model} \cite{Gupta00},
the packet transmitted from node $i$ to node $j$ is successfully delivered when
the distances between node $j$ and the other transmitting nodes are no less than $r$.
If the transmission range $r$ is too large, the transmission often fails
because there are many interfering nodes.
In order to avoid excessive interference, we set the transmission range $r$
to the average distance to the nearest node in the area:
\begin{eqnarray}\label{nearest_node_dist}
r=\int_{0}^{\sqrt{\frac{S}{\pi}}}\left(1-\frac{\pi x^2}{S}\right)^{n-1} dx=\frac{\sqrt{S}\Gamma (n)}{2\Gamma \left(n+\frac{1}{2}\right)}\approx\frac{\sqrt{S}}{2\sqrt{n}},
\end{eqnarray}
where $\Gamma (z)=\int_{0}^{\infty} t^{z-1}e^{-t} dt$ is the gamma function.

\subsection{Two-Phase Routing}

A pair of source and destination nodes is given randomly.
Unless there is the corresponding destination node of a source node in transmission range $r$,
its packet should be delivered via a relay node.
In this paper, the transmission policy follows the two-phase routing \cite{Grossglauser02}:
\begin{itemize}
\item {\it Mode switch.} In each time slot, a node becomes a transmitter or a receiver with probability $q$ or $1-q$, respectively.
\item {\it Phase 1.} In odd time slots, let us consider node $i$ becomes a transmitter.
    If there is at least one receiver within transmission range $r$, node $i$ forwards its packet to one of them.
    This receiver node can be the destination of node $i$.
\item {\it Phase 2.} In even time slots, let us consider node $i$ becomes a receiver.
    If there is at least one transmitter within transmission range $r$ and one of them has a packet whose destination is node $i$,
    it forwards the packet to node $i$.
    This transmitter can be the source of node $i$.
\end{itemize}
In \cite{Grossglauser02}, the throughput of the two-phase routing is defined as follows:
\vskip 10pt
\noindent {\bf Definition 1.} 
(Throughput) {\it Let $M_{i}(t)$ be the number of node $i$ packets
that its corresponding destination node receives during $t$ time slots. We say that a long-term per node throughput of $\Lambda$ is feasible for every S-D pair if:
\begin{eqnarray}\label{Throughput_definition}
\liminf_{t\rightarrow \infty} \frac{1}{t}
M_{i}\left(t\right)\geq \Lambda.
\end{eqnarray}
}
\noindent Hereafter, the long-term per node throughput is abbreviated as the throughput.
When a transmitter forwards a packet, a constant amount of energy is consumed\footnote{{It is implicitly assumed that a modulation and coding scheme (MCS) is fixed and constant power is required to deliver a packet within the transmission range. It is interesting to adjust a control to improve the energy efficiency, which is outside the scope of current work.}}.
In \cite{Ko2013}, it is defined as one {\it unit of energy}. A node is {\it active}
when it has at least one unit of energy. Otherwise, the node is {\it inactive}.
We define the active probability $p_{\mathrm{on}}$ as the probability a node has at least one unit of energy.
In \cite{Ko2013}, the throughput $\Lambda$ is expressed in terms of the active probability $P_{\mathrm{on}}$ as follows:
\begin{eqnarray}\label{Lemma1}
\Lambda=\frac{1}{2}\cdot q\cdot p_{\mathrm{on}}\cdot e^{-\frac{\pi}{4}q p_{\mathrm{on}}}\cdot \left(1-e^{\frac{\pi}{4}(-1+q)}\right).
\end{eqnarray}
It is shown that the throughput $\Lambda$ of the two-phase routing depends on 
the number of active nodes, which is determined by an energy recharging process 
according to our recharging mechanism introduced in the following subsection. 

\subsection{Recharging Mechanism by Wireless Charging Stations}

Inactive nodes cannot transmit the packets in their buffers.
In order to recharge the batteries of them, $m$ WCSs are deployed in the network.
WCSs recharges nodes via magnetic resonance coupling. 
{No interference between data transmission and energy transfer exists because each of them use a separated band. }
 
{
The energy transferred to a mobile is determined by the product of the maximum deliverable units of energy 
$E$ and the energy transfer efficiency $\tau(x)$, where $x$ is the distance to its associated WCS.
Let  $R_y$ denote the maximum distance that a node can receive $y$ units of energy. 
Without loss of generality, the efficiency $\tau(x)$ is a monotone decreasing function of $x$, and the charging range $R_y$ 
is determined by finding the value of $x$ that $E \cdot \tau(x)$ becomes $y$, such that $R_y=\{x:E \cdot \tau(x)=y\}$. 
For tractable analysis, the efficiency  $\tau(x)$ is independent of node speed. 
Let us denote by $Y_j(t)$ the location of WCS $j$ in time $t$. 
The distance 
between node $i$ and WCS $j$ becomes $\| X_i(t)-Y_j(t)\|$ where $\|\cdot\|$ means the Euclidean distance,
and the recharged units of energy $\upsilon\left(\| X_i(t)-Y_j(t)\|\right)$~is
\begin{align}\label{v_x}
&\upsilon(\| X_i(t)-Y_j(t)\|)=\nonumber\\
&\left\{\begin{array}{ll}
E \quad \textrm{if $\| X_i(t)-Y_j(t)\|\leq R_{E}$}\\
k \quad \textrm{if  $R_{k+1} < \| X_i(t)-Y_j(t)\|\leq R_{k}$, $k=1,\cdots,E-1$,}\\
0 \quad \textrm{else,}
\end{array}\right.
\end{align}
where $R_E<R_{E-1}<\cdots<R_1$. 
Throughout this paper, we use the energy transfer efficiency function in \cite{Xie2012_2}, i.e, 
$\tau(x)=-0.0958x^2-0.0377x+1.0$, which is obtained through the curve fitting of the experimental results of \cite{Kurs2010}.
Let us define {\it charging range} as the maximum distance that a node receives at least one unit of energy from a WCS.
Given the recharging mechanism of \eqref{v_x}, the charging range is $R_1$.}
The time required to receive energy from a WCS to a node is extremely short compared to one time slot\footnote{It is a reasonable setting because the maximum power transfer rate of magnetic resonance coupling  is $12$ Watt \cite{Witricity} 
whereas that of an LTE mobile is $23$ dBm.}.
This means that the contact duration is long enough to deliver up to $E$ units of energy 
unless node speed becomes infinite.      

The battery of each node is recharged by one of WCSs.  
{Even though a node is in the charging regions in multiple WCSs,   
it is assumed to associate with only one WCS due to the practical alignment technique, and the maximum recharged energy within one time slot is $E$ units of energy.} 
The maximum battery capacity of each node is set to $L$ units of energy.
If the sum of residual and recharged units of energy are larger than $L$,
a node saves $L$ units of energy only, and the remaining is thrown out.
Each WCS can recharge up to $u$ nodes at a time\footnote{{The number $u$ depends on the technique to track the resonance frequency. For example, it is experimentally shown in \cite{Cannon2009} and \cite{Fu2015} that up to two devices can be charged by using the technique of the said resonant frequency splitting and load balancing, respectively. }}.
When there are more than $u$ nodes within the charging region,
WCS randomly selects $u$ nodes among them.

Each WCS always monitors own remaining energy.
If the remaining energy is below a certain level,
it communicates with an operator station by using its communication module.
The operator station then sends the charging vehicle, which recharges the WCS before its battery runs out.
This means that all WCSs always have sufficient energy.

 \vspace{10pt}
\section{\uppercase{Stochastic Modeling of Mobile Networks with Wireless Charging Stations}}

In this section, we design a two-dimensional Markov chain in
which the horizontal and vertical state dimensions represent the residual energy and the distance to the nearest WCS, respectively.
We first outline our Markov chain design, and then derive its steady state probabilities to determine the active probability $P_{\mathrm{on}}$ in  \eqref{Lemma1}.

\subsection{Two-Dimensional Markov Chain}
\begin{figure*} [t!]
\centering
\includegraphics[angle=0,width=6.5in]{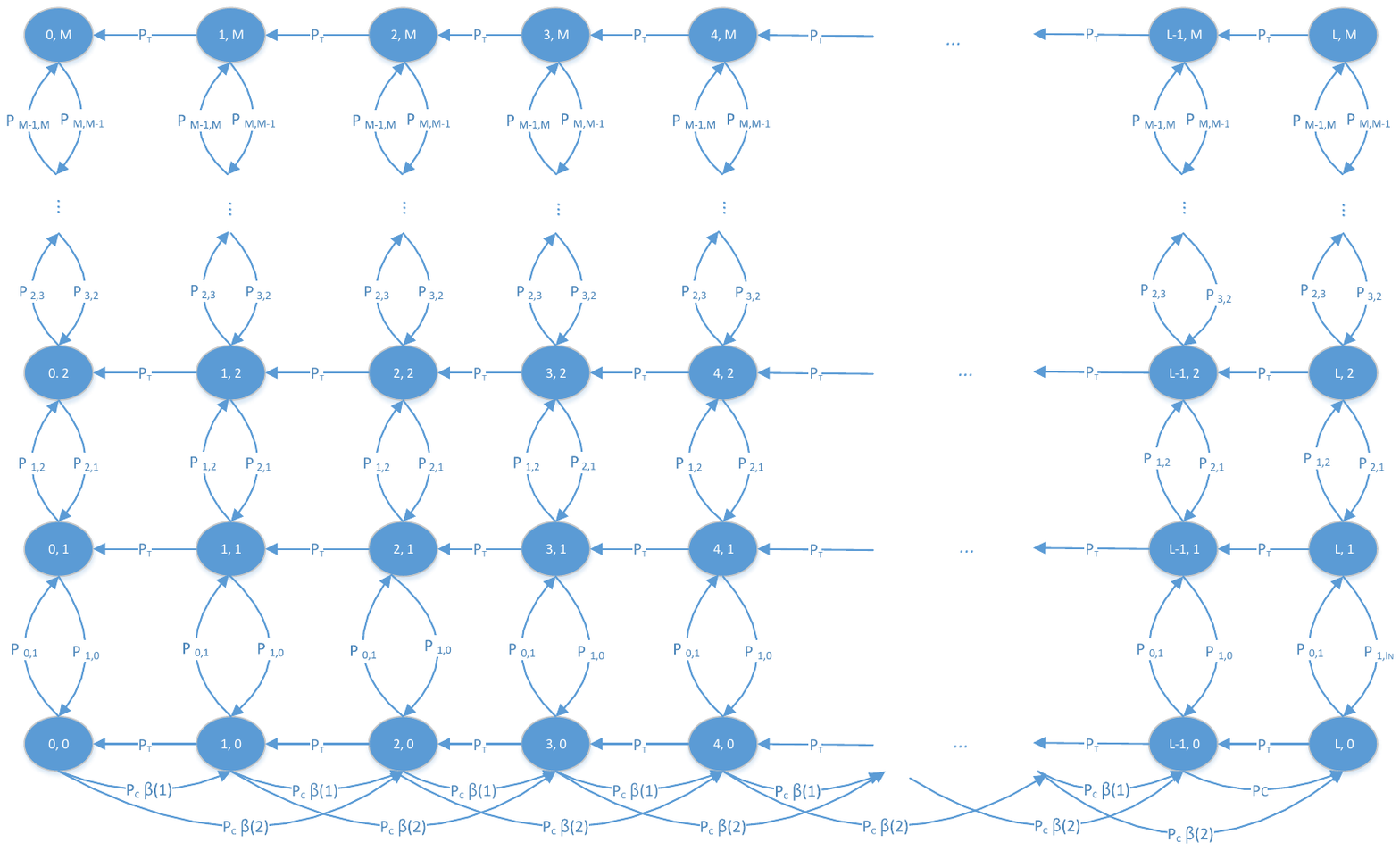}
\caption{Two-dimensional Markov chain of which the horizontal and vertical state dimensions represent
the number of remaining units of energy and the relative distance to the nearest WCS normalized by node speed, respectively.} \label{state_diagram}
\end{figure*}

The state space of our two-dimensional Markov chain $\Psi$ is given as follows:
\begin{eqnarray}\label{State_Space}
\Psi=\left\{\left(e,d\right): 0\leq e\leq L, 0 \leq d\leq M \right\},
\end{eqnarray}
where parameter $e$ is the number of remaining units of energy,
and $d$ is a discrete number indicating the distance to the nearest WCS by the following rule\footnote{{The mobile is assumed to always have at least one packet to transmit. It is possible to figure out random data arrival by plugging one more dimension in the Markov chain model in Fig. \ref{state_diagram}, which remains as a future work.}}:
\setlength\arraycolsep{0.1em}\begin{eqnarray}\label{partition}
d=\left\{\begin{array}{ll}
0& \textrm{if $\min_{j}\| X_i(t)-Y_j(t)\|\leq R_1$,}\\
1& \textrm{else if $\min_{j}\| X_i(t)-Y_j(t)\|\leq R_1+v$,}\\
\qquad\qquad \vdots & \textrm{}\\
k & \textrm{else if $\min_{j}\| X_i(t)-Y_j(t)\|\leq R_1+kv$,}\\
\qquad\qquad \vdots & \textrm{}\\
M & \textrm{otherwise.}\\
\end{array}\right.
\end{eqnarray}
The number $M$ in \eqref{partition} is interpreted as the resolution of the Markov chain
in the sense that larger $M$ can express the movement pattern of a node more accurately.
The number $d$ is understood as a {\it relative distance}
at the point that a physical distance is normalized by node speed $v$.

Figure \ref{state_diagram} represents an example of the two-dimensional Markov chain
when WCS can deliver up to two units of energy to a node within one time slot ($E=2$).
There are threefold state transitions as follows:

\begin{itemize}
\item The state transition to the up or down arises
when the relative distance to the nearest WCS $d$ of \eqref{partition} becomes further or closer, respectively. 
Let $P_{i,j}$ detnote the probability that the relative distance $d$ is changed from $i$ to $j$, i.e, 
\setlength\arraycolsep{0.1em}\begin{eqnarray}\label{Pij}
P_{i,j}(t)=\Pr\left[\textrm{$d=j$ at ${t+1}$ slot}|\textrm{$d=i$ at $t$ slot}\right].
\end{eqnarray}
The mobility model follows a time-invariant Markovian process of 
which the transition probabilities are constant regardless of  slot $t$, and 
$P_{i,j}(t)$ can be simply expressed as $P_{i,j}$ by  omitting the index $t$. 
The exact form of $P_{i,j}$ and its derivation process are in Appendix~A.
It is worth mentioning 
that all transition probabilities $P_{i,j}$ are constant regardless of the residual energy status.
A node cannot move to the charging region intentionally because it does not know WCS's location.
Its energy status is thus determined by {\it incident} contacts to other nodes or WCSs.     
\item The state transition to the left arises 
when the node transmits a packet to one of neighbours nodes.
Let $p_{t}$ denote a probability that an active node can transmit its packet as
\setlength\arraycolsep{0.1em}\begin{eqnarray}
p_t&=& q\cdot\left[1-\left\{1-(1-q)\frac{\pi {r}^2}{S}\right\}^{n-1} \right], \label{p_t}
\end{eqnarray}
of which the detailed derivation process is in Appendix B.
Unless its residual energy $e$ is zero, 
the transmission probability $p_{c}$ is constant regardless of the relative distance $d$ of \eqref{partition}.
\item The state transition to the right arises when the node is recharged by a WCS.
This event only happens when the node is selected by one of WCSs is in the charging region, and
these are only stipulated on the lowest state transition ($d=0$). 
Recall that each WCS can charge up to $u$ nodes in a given  slot.  
We define a charging probability $p_{c}$ as the possibility the node becomes one of $u$ selected nodes, i.e,
\setlength\arraycolsep{0.1em}\begin{align}
&p_c=\frac{1-\gamma(u,n)^m}
{1-\left(1-\frac{\pi {R_1}^2}{S}\right)^m}, \label{p_c}
\end{align}
where
\begin{align} 
\gamma(u,n)=&1-\frac{\pi {R_1}^2}{S} F(u-2;n-1,\frac{\pi {R_1}^2}{S})\nonumber\\
&-\frac{u}{n}\left(1-F(u-1;n,\frac{\pi {R_1}^2}{S}\right),
\end{align}
and $F(k;n,p)=\sum_{i=0}^k {n \choose i} p^i (1-p)^{n-i}$ is the cumulative distribution function (CDF) of the binomial distribution with parameters $k$, $n$ and $p$. Its derivation process is in Appendix B.  
The number of  recharged units of energy depends on the distance to its associated WCS.
Let $\beta(i)$ denote a probability a node receives $i$ units of energy as follows:       
\setlength\arraycolsep{0.1em}\begin{eqnarray}\label{beta}
\beta(i)=
\left\{\begin{array}{ll}
\frac{{R_i}^2-{R_{i+1}}^2}{{R_1}^2} & \quad \textrm{if $i=1,\cdots, E-1$,}\\
\frac{{R_i}^2}{{R_1}^2} &\quad \textrm{if $i=E$.}\\
\end{array}\right.\nonumber
\end{eqnarray}
A node in the charging region thus receives $i$ units of energy with probability $p_{c} \beta(i)$.
\end{itemize}
\subsection{Steady State Probability and Throughput}

Let $\pi_{e,d}$ denote the steady state probability when the residual energy is $e$ and the relative distance is $d$.
Then, we make the following steady state vector $\boldsymbol{\pi}
=\left[\begin{array}{ccccccccc}
\pi_{0,0}, \cdots  \pi_{0,M}, & \pi_{1,0}, \cdots  \pi_{1,M},  \cdots &
\pi_{L,0}, \cdots  \pi_{L,M}\\
\end{array}\right]$,
which is partitioned according to the number of remaining units of energy, i.e,
\setlength\arraycolsep{0.1em}\begin{eqnarray}\label{partition_steady_state_probability}
\boldsymbol{\pi}=\left[\begin{array}{cccc}
\boldsymbol{\pi_0} &
\boldsymbol{\pi_1} &
\cdots &
\boldsymbol{\pi_L}\\
\end{array}\right],
\end{eqnarray}
where
\setlength\arraycolsep{0.1em}\begin{eqnarray}\label{each_partition_steady_state_probability}
&&\boldsymbol{\pi_i}=\left[\begin{array}{cccc}
\pi_{i,0} &
\pi_{i,1} &
\cdots &
\pi_{i,M}\\
\end{array}\right].
\end{eqnarray}
In order to derive $\boldsymbol{\pi}$, we make the following balance equation:
{\setlength\arraycolsep{0.1em}\begin{eqnarray}\label{Equation_steady_state_probability}
\boldsymbol{\pi}\boldsymbol{Q}=0, \quad
\boldsymbol{\pi}\boldsymbol{1}=1,
\end{eqnarray}
where $\boldsymbol{1}$ is the column vector where every entity is one,
and $\boldsymbol{Q}$ is the generating matrix of the corresponding Markov chain:
\setlength\arraycolsep{0.1em}
\medmuskip=-1mu
\thinmuskip=-1mu
\begin{eqnarray}\label{Q}
\boldsymbol{Q}=
\left(\begin{array}{ccccccccc}
\boldsymbol{B_0} & \boldsymbol{A_2} & \boldsymbol{A_3} & 0 & 0      & \dots & 0   & 0   & 0\\
\boldsymbol{A_0} & \boldsymbol{A_1} & \boldsymbol{A_2} & \boldsymbol{A_3} & 0    & \dots & 0   & 0   & 0\\
0   & \boldsymbol{A_0} & \boldsymbol{A_1} & \boldsymbol{A_2} & \boldsymbol{A_3}    & \dots & 0   & 0   & 0\\
\vdots & \vdots & \vdots & \vdots & \vdots  & \ddots & \vdots & \vdots\\
0   & 0   & 0   & 0   & 0      & \dots & \boldsymbol{A_1} & \boldsymbol{A_2} & \boldsymbol{A_3}\\
0   & 0   & 0   & 0   & 0      & \dots & \boldsymbol{A_0} & \boldsymbol{A_1} & \boldsymbol{A_2}+\boldsymbol{A_3}\\
0   & 0   & 0   & 0   & 0      & \dots & 0   & \boldsymbol{A_0} & \boldsymbol{A_1}+\boldsymbol{A_2}+\boldsymbol{A_3}\\
\end{array}\right).
\end{eqnarray}
Its sub-matrices $\boldsymbol{B_0}$, $\boldsymbol{A_0}$, $\boldsymbol{A_1}$, $\boldsymbol{A_2}$ and 
$\boldsymbol{A_3}$ are expressed as follows:
\setlength\arraycolsep{0.1em}
\medmuskip=-1mu
\thinmuskip=-1mu
\begin{eqnarray}\label{B0}
\mathbf{B_0}
=
\left(\begin{array}{ccccc}
-P_{0,1}-p_c & P_{0,1}                  & 0                  & \cdots   & 0\\
P_{1,0}      &-P_{1, 0}-P_{1,2} & P_{1,2}              & \cdots   & 0\\
0                     & P_{2,1}                  &-P_{2,1}-P_{2,3} &  \cdots    & 0\\
\vdots                     & \vdots                 &\vdots  & \ddots &     \vdots\\
0                     & 0                       & 0                  & \cdots  & -P_{M, M-1}\\
\end{array}\right),\nonumber
\end{eqnarray}

{\setlength\arraycolsep{0.1em}\begin{eqnarray}\label{A0A1}
\mathbf{A_0}
=\left(\begin{array}{ccc}
p_t                 & \cdots    & 0\\

\vdots    & \ddots    & \vdots\\
0                  & \cdots  & p_t\\
\end{array}\right)=p_t\mathbf{I},\quad
\mathbf{A_1}=\mathbf{B_0}-\mathbf{A_0},\nonumber
\end{eqnarray}


\setlength\arraycolsep{0.1em}\begin{eqnarray}\label{A2A3}
\medmuskip=-1mu
\thinmuskip=-1mu
\mathbf{A_2}
=\left(\begin{array}{ccc}
p_c\beta(1)                   & \cdots    & 0\\

\vdots                      & \ddots    & \vdots\\
 0                  & \cdots  & 0\\
\end{array}\right),\quad \nonumber
\mathbf{A_3}
=\left(\begin{array}{ccc}
p_c\beta(2)               & \cdots    & 0\\
\vdots                     & \ddots    & \vdots\\
 0                  & \cdots  & 0\\
\end{array}\right).\nonumber
\end{eqnarray}
After solving the balance equation of \eqref{Equation_steady_state_probability},
we acquire the active probability $P_{\mathrm{on}}$ as 
{\setlength\arraycolsep{0.1em}\begin{eqnarray}\label{Active_Probability}
P_{\mathrm{on}}=\sum_{i=1}^L \boldsymbol{\pi_i} \boldsymbol{1}=1-\boldsymbol{\pi_0} \boldsymbol{1}.
\end{eqnarray}
With \eqref{Lemma1}, the throughput $\Lambda$ is given as
{\setlength\arraycolsep{0.1em}\begin{eqnarray}\label{Throughput_from_steady_state_probabilitys}
\Lambda
=\frac{1}{2}\cdot q\cdot \left(1-\boldsymbol{\pi_0} \boldsymbol{1}\right)\cdot e^{-\frac{\pi}{4}q \left(1-\boldsymbol{\pi_0} \boldsymbol{1}\right)}\cdot \left(1-e^{\frac{\pi}{4}(-1+q)}\right).
\end{eqnarray}

\section{Performance Evaluation of Mobile Ad-Hoc Networks\\ with Wireless Charging Stations}

\subsection{Inter-meeting time and Throughput}\label{subsection_throughput_1}

\begin{figure*} [t!]
\centering
\subfigure[Inter-meeting time]{\includegraphics[angle=0,width=3.50in]{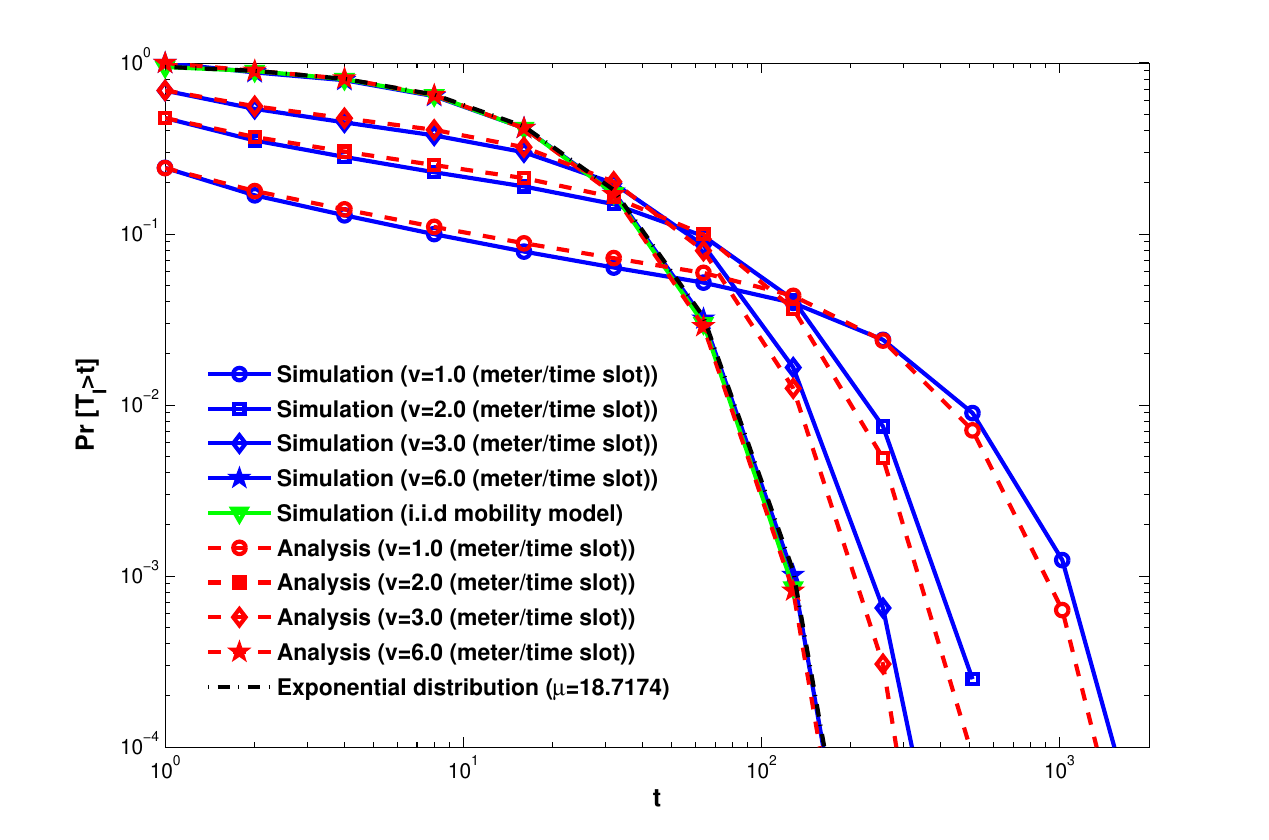}}
\subfigure[Throughput]{\includegraphics[angle=0,width=3.50in]{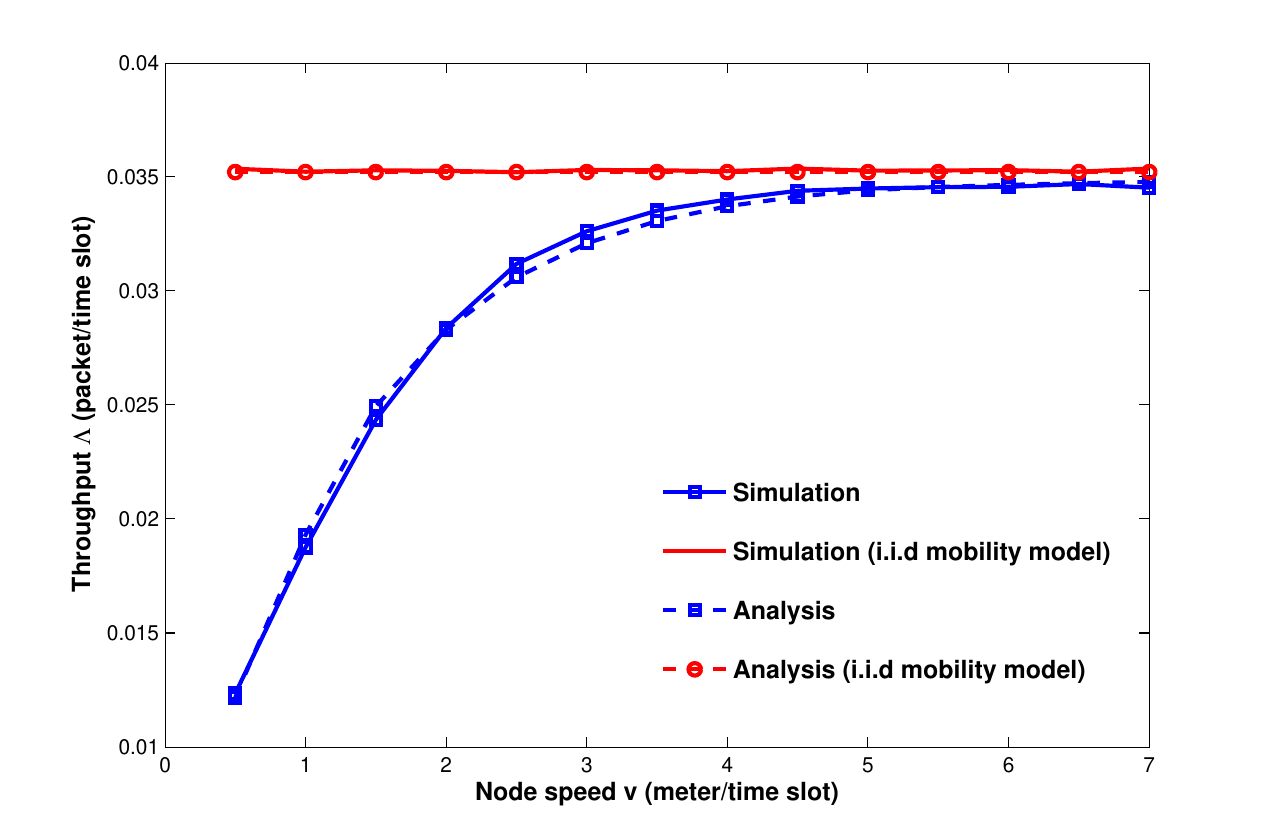}}
\caption{(a) The stochastic distribution of inter-meeting time under various speed $v$. (b) Throughput as a function of speed $v$ ($S=20$, $L=10$, $q=0.5$, $u=1$, $E=3$, $n=10$, $m=1$).} \label{Throughput_versus_speed}
\end{figure*}

In this subsection, we explain how node speed $v$ affects the throughput $\Lambda$ by means of inter-meeting time defined as follows:
\vskip 10pt

\noindent {\bf Definition 2.} (Inter-meeting time)
{\it Consider there are node $i$ and WCS $j$ in the network.
The inter-meeting time $T_I$ is the interval between adjacent meeting events between node $i$ and WCS $j$.
\setlength\arraycolsep{0.1em}
\medmuskip=-1mu
\thinmuskip=-1mu
\begin{eqnarray}\label{T_I}
T_I=\inf\left\{t\geq0:Z_{t+k}=1\mid Z_{k}=1\right\},
\end{eqnarray}
where $Z_{t}$ is an indicator to check whether a meeting event occurs between node $i$ and WCS $j$ at time $t$.
If $\| X_i(t)-Y_j(t)\|\leq R_1$, we set $Z_{t}$ to one.
Otherwise,  $Z_{t}=0$.}
\vskip 10pt
{The inter-meeting time $T_{I}$ is closely linked to an energy starving period 
because a node has no opportunity to receive energy until one of WCSs is encountered.}
The stochastic features of $T_{I}$ is thus related to
an energy provision process of an arbitrary node. 
Let us denote by $\boldsymbol{P}$ an $M$ by $M$ matrix
of which the elements represents the transition probability $P_{i,j}$ \eqref{Pij} ($1\leq i,$ $j\leq M$):
{\setlength\arraycolsep{0.1em}\begin{eqnarray}\label{P}
\boldsymbol{P}=\left(\begin{array}{c}
\boldsymbol{p_1}\\
\boldsymbol{p_2}\\
\vdots  \\
\boldsymbol{p_M}
\end{array}\right)
=\left(\begin{array}{ccccc}
P_{1,1} & P_{1,2} & \dots & P_{1,M-1} & P_{1,M}\\
P_{2,1} & P_{2,2} & \dots & P_{1,M-1} & P_{2,M}\\
\vdots  & \vdots  & \ddots& \vdots    & \vdots\\
P_{M,1} & P_{M,2} & \dots & P_{M-1,M} & P_{M,M}
\end{array}\right),
\end{eqnarray}
where
{\setlength\arraycolsep{0.1em}\begin{eqnarray}
\boldsymbol{p_i}=\left(\begin{array}{ccccc}
P_{i,1} & P_{i,2} & \dots & P_{i,M-1} & P_{i,M}\\
\end{array}\right).
\end{eqnarray}
From $\boldsymbol{P}$ \eqref{P}, we derive the stochastic distribution of inter-meeting time $T_I$ in the following Proposition:

\vskip 10pt
\noindent {\bf Proposition 1.} {\it The complementary cumulative distribution function (CCDF) of the inter-meeting time $T_I$ is
 \begin{eqnarray}
\Pr\left[T_I>t\right]=\sum_{i=1}^{M}\gamma_i  \lambda_i^{t-1}\label{Proposition_1},
\end{eqnarray}
where $\lambda_i$ is the $i^{th}$ eigenvalue of $\boldsymbol{P}$ (\ref{P}) ($1> \lambda_1 > \cdots > \lambda_M > 0$).
The coefficient $\gamma_i$ is
\begin{eqnarray}\label{gamma}
\gamma_i&=&\boldsymbol{p_0} \boldsymbol{a_i} \boldsymbol{b_i^T},\nonumber
\end{eqnarray}
where
vectors $\boldsymbol{a_i}$ and $\boldsymbol{b_i}$ are the right-hand and left-hand eigenvectors of $\lambda_i$ such that
$\boldsymbol{P}\boldsymbol{a_i}=\lambda_i \boldsymbol{a_i}$ and $\boldsymbol{b_i^*}\boldsymbol{P}=\lambda_i \boldsymbol{b_i^*}$\footnote{$\boldsymbol{x}^*$ is a conjugate transpose of $\boldsymbol{x}$.}, respectively.
}
\vskip 10pt

\noindent {\bf Proof.}
See Appendix D.
\hfill $\blacksquare$
\vskip 10pt

Figure \ref{Throughput_versus_speed} (a) depicts the CCDFs of inter-meeting time $T_I$.
We numerically measure the inter-meeting time $T_I$  by changing the node speed as $v=1.0,$ $2.0$, $3.0$ and $6.0$ (meters/slot).
The higher is the node speed $v$, the less frequent are lengthy inter-meeting times.
A node with a higher speed can reach the charging region of the WCS within a few time slots, reducing the occurrence of lengthy inter-meeting times.
A node with a higher speed can move farther from its previous location,
and whether or not to encounter a WCS solely depends on the ratio of the charging region to the network area, i.e., $\frac{1}{\mu}=\frac{\pi {R_1}^2}{S^2}\approx 0.053$ as does the i.i.d. mobility model.
With increased node speed,  
the distribution converges to that of the i.i.d. mobility model following the exponential distribution with parameter
${\mu}\approx18.7174$.

The CCDF of $T_I$ of \eqref{Proposition_1} is the  sum of powered eigenvalues with the exponent $t$. 
As $t$ becomes larger,  we approximate it in terms of the largest eigenvalue $\lambda_1$ because other terms decay faster:
\begin{eqnarray}\label{Intermeeting_distribution_approximation}
\Pr\left[T_I>t\right]\approx \lambda_1^{t}.
\end{eqnarray}
The eigenvalue $\lambda_{1}$ is called the {\it spectral radius} of matrix $\boldsymbol{P}$ \eqref{P}.
As the spectral radius becomes smaller, 
the approximated CCDF \eqref{Intermeeting_distribution_approximation} decreases more sharply
especially when $t$ is large. 
This indicates that lengthy inter-meeting times are infrequent when $\lambda_{1}$ is small. 
In Table \ref{Table_2}, we summarize this spectral radius $\lambda_1$ as a function of node speed $v$ 
and show that $\lambda_1$ is a non-increasing function of node speed $v$.
Consequently,  a higher node speed decreases spectral radius $\lambda_{1}$ and produces fewer occurrences of lengthy inter-meeting times.

\begin{table*}[t]
  \centering
  \begin{footnotesize}
  \begin{tabular}{|c|c|c|c|c|c|c|c|c|c|c|c|c|c|}
  \hline
   & v=0.5 & v=1.0 & v=1.5 & v=2.0 & v=2.5 & v=3.0 & v=3.5 & v=4.0 & v=4.5 & v=5.0 & v=5.5 & v=6.0\\
  \hline
  $\lambda_1$  & 0.9985  & 0.9953  & 0.9903 & 0.9845 & 0.9780 & 0.9714 & 0.9649 & 0.9585 & 0.9534 & 0.9492 & 0.9471 & 0.9457 \\
  \hline
\end{tabular}
\end{footnotesize}
\caption{Spectral radius $\lambda_1$ as a function of node speed $v$.}\label{Table_2}
\end{table*}

The above feature of the inter-meeting time affects the energy provision process.
Figure \ref{Throughput_versus_speed}~(b) shows this impact.
When node speed $v$ is $0.5$ (meter/time slot), the throughput is nearly one-third of that of the i.i.d. mobility model.
A node is unable to receive energy for a long time due to the lengthy inter-meeting time and remains in an inactive state.
This results in a decrease in throughput.
As $v$ increases, on the other hand, the inter-meeting time decreases.
This leads to a reduction in energy-starving period and improvement of throughput.

\subsection{Battery Capacity and Throughput}

Consider a slow-moving node with a rather longer sojourn time, the duration
a node remains in the charging region. 
The node can receive energy continuously from the associated WCS.
Nevertheless, the node is unable to save more than $L$ units of energy due to the battery capacity constraint.
In other words, the node can remain active longer  if the battery capacity were to be increased. 
We start with the following proposition.

\vskip 10pt
\noindent {\bf Proposition 2.} {\it When the battery capacity $L$ becomes infinite, the throughput
of an energy-constrained network $\Lambda$ is
\setlength\arraycolsep{0.1em}
\medmuskip=-2mu
\thinmuskip=-2mu
\begin{eqnarray}\label{Proposition_2}
\Lambda
&=&\frac{q}{2} \min\left[1, \frac{p_c}{p_t}\left\{1-\left(1-\frac{\pi {R_1}^2}{S}\right)^m\right\}\left(\sum_{k=1}^E k \beta(k)\right)\right]\left(1-e^{\frac{\pi(q-1)}{4}}\right) \nonumber\\
&&\cdot e^{-\frac{\pi}{4}q\cdot
\min\left[1,\frac{p_c}{p_t}\left\{1-\left(1-\frac{\pi {R_1}^2}{S}\right)^m\right\}\left(\sum_{i=k}^E k \beta(k)\right)\right]},
\end{eqnarray}
which is independent of node speed $v$.
}

\vskip 10pt
\noindent {\bf Proof.}
See Appendix E. 
\hfill $\blacksquare$
\vskip 10pt

\begin{figure} [t!]
\centering
\includegraphics[angle=0,width=3.5in]{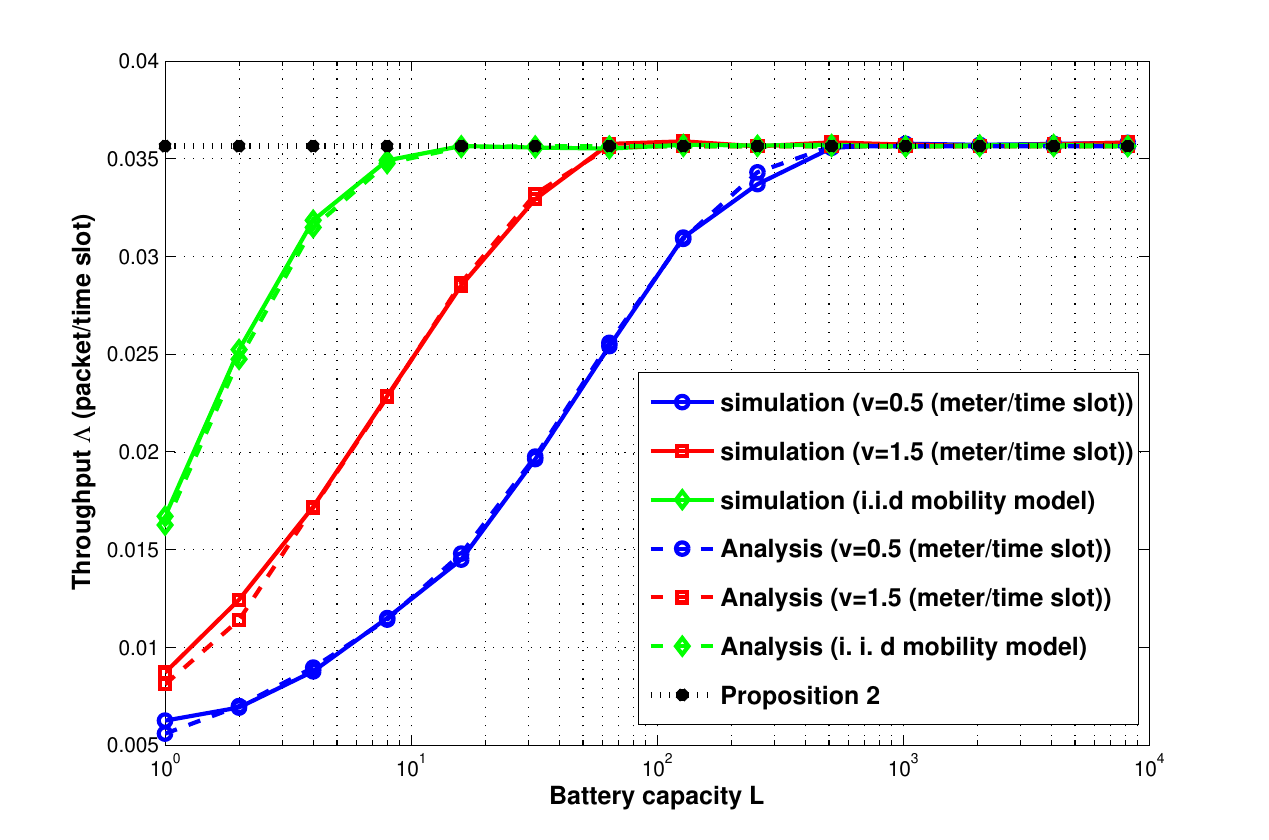}
\caption{Throughput as a function of battery capacity $L$ ($S=20$, $q=0.5$, $u=1$, $E=3$, $n=10$, $m=1$).} \label{Throughput_versus_battery}
\end{figure}

Figure \ref{Throughput_versus_battery} represents the throughput $\Lambda$ as a function of battery capacity $L$.
As the battery capacity $L$ increases,
the throughput increases and converges according to Proposition 2 \eqref{Proposition_2} (see the black dotted line).
An interesting point is that Proposition~2 is achievable even under a finite battery  capacity.
If a node can store enough energy to sustain the inter-meeting time,
it remains in an active state and achieves the throughput in Proposition 2.
We calculate the mean of the inter-meeting time $E[T_{I}]$ 
utilizing Equation \eqref{Intermeeting_distribution_approximation} and the spectral radius $\lambda_1$ in Table \ref{Table_2}.
\begin{eqnarray}
E[T_I]=\sum_{t=0}^\infty \Pr\left[T_I>t\right]
\approx\sum_{t=0}^\infty {\lambda_1}^t=\frac{1}{1-\lambda_1}
\end{eqnarray}
When the battery capacity $L$ is no less than $E[T_{I}]$,
the throughput $\Lambda$ becomes the same as that in Proposition 2 \eqref{Proposition_2}.
For example, when node speed $v$ is $0.5$ or $1.5$ (meter/slot),
its spectral radius $\lambda_{1}$ is $0.9985$ or $0.9903$ (see Table \ref{Table_2})
and its corresponding $E[T_{I}]$ becomes $666.67$ or $103.09$, respectively.
As a result, a battery capacity larger than  $E[T_{I}]$ is a necessary condition to achieve Proposition 2. 

\subsection{Node Density and Throughput}

Since the seminal work by Grossglauser and Tse \cite{Grossglauser02}, 
investigating the relationship between throughput $\Lambda$ and node density $n$ has been the most fundamental issue with mobile networks; 
therefore yet the impact of irregular energy provision due to low node speed has not been studied.
In this subsection, we investigate this effect through  some numerical evaluations and the following throughput scaling law. 
 
 \vskip 10pt
\noindent {\bf Proposition  3.} {\it
The scaling law of the throughput $\Lambda$ is:
\begin{eqnarray}\label{throughput_scaling_law}
\Lambda=
\Theta\left(\min\left(1,\frac{m}{n}\right)c^{\min\left(1,\frac{m}{n}\right)}\right),
\end{eqnarray}
where $0<e^{-\frac{\pi \cdot u}{4\cdot a}}<c\leq e^{-\frac{\pi \cdot u}{4\cdot a}\left(\sum_{k=1}^{E}k\beta(k)\right)}<1$, and $a=1-e^{-\frac{\pi}{4}(1-q)}$.
}
\vskip 10pt
\noindent {\bf Proof.}
See Appendix F. 
\hfill $\blacksquare$
\vskip 10pt
Proposition 3 indicates that the throughput $\Lambda$ is a function of the ratio of the number of WCSs $m$ 
and the number of nodes $n$, 
and independent of node speed $v$.  
A node with low speed receives energy from WCSs irregularly, yielding the decrease of throughput.
Compared with fast-moving one, it needs more WCSs to maintain the same throughput.  
As the network becomes denser, however, the penalty due to slow speed disappears 
and we only consider the ratio $\frac{m}{n}$ when installing WCSs.  
In order to achieve the constant throughput of $\Theta(1)$ as in \cite{Grossglauser02},  for example, 
$\Theta(n)$ WCSs is required regardless of node speed.

\begin{figure*} [t!]
\centering
\subfigure[The number of WCSs $m=1$]{\includegraphics[angle=0,width=3.5in]{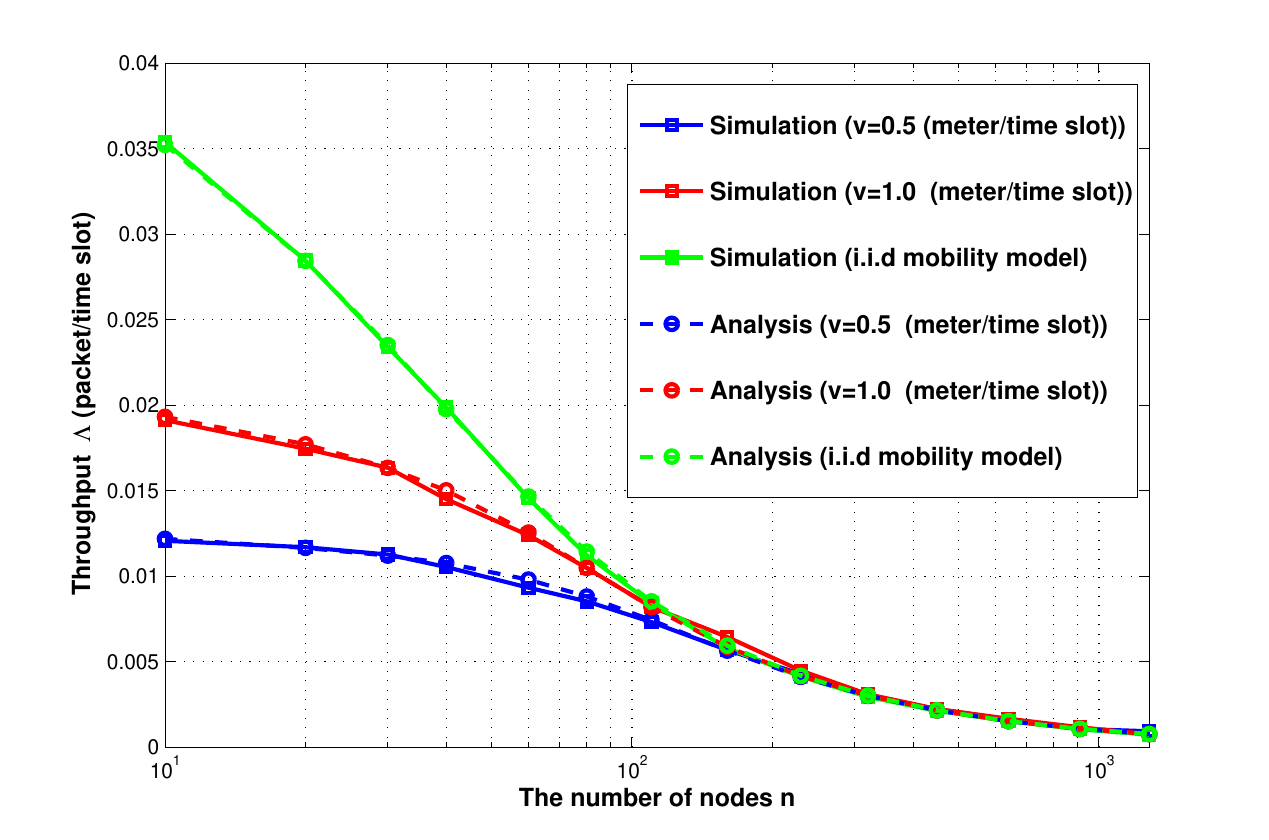}}
\subfigure[The number of WCSs $m=\frac{n}{10}$]{\includegraphics[angle=0,width=3.5in]{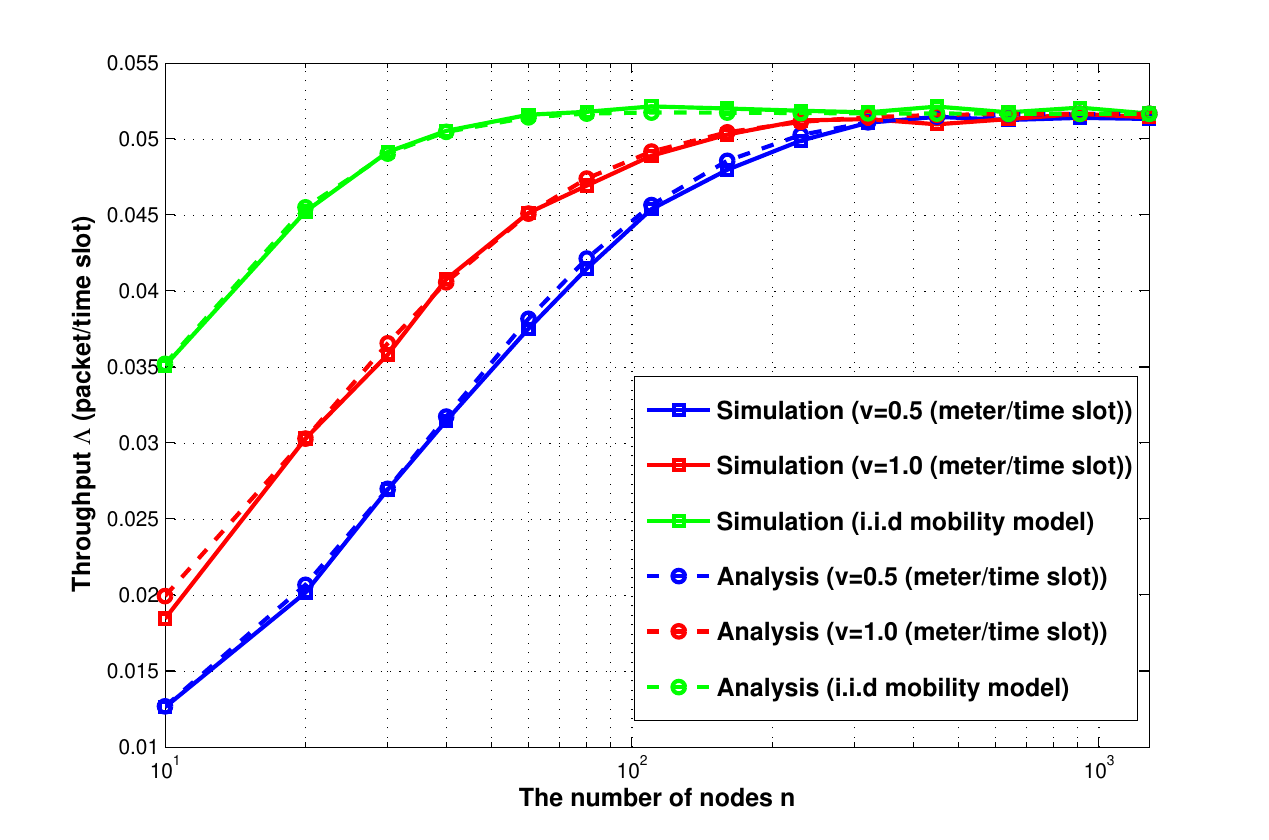}}
\caption{ Throughput as a function of node density $n$ ($S=20$, $q=0.5$, $u=1$, $E=3$, $L=10$).} \label{Throughput_versus_density}
\end{figure*}

{Note that the scaling law in Proposition 3 of \eqref{throughput_scaling_law} is the same as that of the i.i.d. mobility model in \cite{Ko2013}.
Figure \ref{Throughput_versus_density} shows that the throughput $\Lambda$ always converges to that of the i.i.d. mobility model as the number of nodes $n$ increases.}
This implies that a high node density makes nodes look 
as if they are moving at a fast speed in the sense that the i.i.d. mobility model allows a node to increase moving speed $v$ up to the network size. When calculating the throughput of a dense mobile network with WPT,
it is a reasonable assumption that nodes move according to the i.i.d. mobility model.

\section{Concluding Remarks}

In this paper, we determined the impact of node speed on
the throughput of an energy-constrained mobile network
where WPT-enabled apparatuses, known as WCSs,
are deployed and recharge nodes within their charging regions.
There are two distinct energy provision patterns 
according to node speed difference.
A slow-moving node outside a WCS's charging region waits a long time for energy supply from WCSs,
whereas one inside the charging region recharges its battery consistently.
On the other hand, a fast-moving node enables energy to be delivered from a WCS within a short interval.
Such a node receives energy in a regular manner, contrary to the slow-moving node.
The analytic and numerical results showed that this distinct energy-provisioning process yields
a throughput difference between slow- and fast-moving users
when the battery capacity is finite and the network is sparse.
On the other hand, if the battery capacity of a node is large enough to save sufficient energy from WCSs
or the network becomes denser, the difference between the two sets of results disappears.
These findings provide some guidelines for mobile network architectures with WPT such as IoT.
First, the charging opportunity of each node should be prioritized according to moving speed.
Once a slow-moving node leaves a charging region, it will require a long amount of time until it visits the charging region again,
and a WCS should recharge its battery until it is full.
On the other hand, a fast-moving node does not need to charge its battery preferentially because it can re-enter 
the charging region within a short interval.           
Second, installing WCSs in sparse regions with high mobility, such as motorways,
is an efficient energy-provisioning strategy. By exploiting the fast moving speed of vehicles, these WCSs
deliver energy to mobile nodes in a regular pattern.
In dense regions, on the other hand, 
the distinct energy provisioning coming from the node speed difference disappears 
and only the ratio of density between nodes and WCSs determines   
the throughput performance.

A weakness of this study is the simple mobility model where each node moves without preference.
In a real network, on the other hand, users are likely to visit some popular places frequently,
resulting in an energy supply shortage due to the relatively high node density in the area.
Further work should therefore involve user preference and verify its impact on throughput.
Another extension of this work is to inform users of the locations of WCSs.
Moreover, considering the economic aspect of WCSs is another interesting avenue for future research.

\section*{Appendix}

\subsection{Transition Probability $P_{i,j}$ of \eqref{Pij}}\label{Appendix:DerivationPij}

Let $D_t$ denote the distance between a node and a WCS at time $t$.
Since nodes and WCSs are uniformly distributed in a torus area, 
the conditional probability that $D_{t+1}$ is smaller than or equal to $x_2$ given $D_{t}=x_1$ is
\begin{align}\label{conditional_CDF_of_di_t}
&\Pr\left[D_{t+1}\leq x_2 | D_{t}=x_1\right]\nonumber\\
=&\left\{\begin{array}{ll}
0 & \textrm{if $v+x_2<x_1$ or $v-x_2>x_1$,}\\
\frac{\arccos\left(\frac{v^2+{x_1}^2-x_2^2}{2vx_1}\right)}{\pi} & \textrm{if $|v-x_2|\leq x_1<v+x_2$,}\\
1 & \textrm{if $x_2-v>x_1$,}\\
\end{array}\right.
\end{align}
which is based on the fact that nodes and WCSs are uniformly distributed in a torus area. 
From the conditional probability \eqref{conditional_CDF_of_di_t},
we derive the joint cumulative distribution function (CDF) that $D_t$ is smaller than or equal to $x_1$, and $D_{t+1}$  is smaller than or equal to $x_1$:
\begin{align}\small\label{joint CDF_of_di_t}
&\Pr\left[D_t\leq x_1, D_{t+1}\leq x_2\right]\nonumber\\
=&\int_0^{x_1} \Pr\left[D_{t+1}\leq x_2 | D_{t}=x\right]f_{D_{t}}(x)dx.
\end{align}
Using  \eqref{joint CDF_of_di_t}, we calculate the following joint probability:
{\setlength\arraycolsep{0.1em}
\begin{align}\label{joint probability_of_di_t}
&\Pr\left[x_1 \leq D_{t}\leq x_2,  x_3 \leq D_{t+1}\leq x_4\right]\nonumber\\
=&\Pr\left[D_t\leq x_2, D_{t+1}\leq x_4\right]-\left[D_t\leq x_1,  D_{t+1}\leq x_4\right]\nonumber\\
&-\Pr\left[D_t\leq x_2,  D_{t+1}\leq x_3\right]+\Pr\left[D_t\leq x_1,  D_{t+1}\leq x_3\right].
\end{align}
By inserting the boundary values of the $i$ and $j$ states in \eqref{partition} into $x_1$, $x_2$, $x_3$ and $x_4$ in \eqref{joint probability_of_di_t},
we can derive the joint probability $\alpha_{i,j}$ that relative distances $d$ \eqref{partition} at time slot $t$ and $t+1$ are $i$ and $j$, respectively.  
In order to calculate $\alpha_{0,1}$, for example, we set $x_1=0$, $x_2=R_1$, $x_3=R_1$ and $x_4=R_1+v$.

Define $A_{a,b}$ as the joint CCDF that both relative distances $d_t$ and $d_{t+1}$ are equal to or larger than $a$ and $b$ when the number of WCSs $m$ is one, which is expressed as the sum of $\alpha_{i,j}$, i.e.,
\begin{align}\label{Aab}
A_{a,b}
&=\sum_{i=a}^M\sum_{j=b}^M\alpha_{i,j}.
\end{align}
Noting that each location of WCSs is independent, we derive $P_{i,j}$ in terms of $A_{a,b}$ as follows:
\begin{itemize}
\item If $i=0$,
\setlength\arraycolsep{0.1em}
\medmuskip=-1mu
\thinmuskip=-1mu
\begin{eqnarray}\label{Pij_1_proof}
P_{0,j}=\left\{\begin{array}{ll}
1-\frac{{A_{0,1}}^m-{A_{1,1}}^m}
{1-\left(1-\frac{\pi {R_1}^2}{S}\right)^m} & \quad\quad\quad\quad\quad\quad\quad\textrm{if $j=0$,}\\
\frac{{A_{0,1}}^m-{A_{1,1}}^m}
{1-\left(1-\frac{\pi {R_1}^2}{S}\right)^m} & \quad\quad\quad\quad\quad\quad\quad\textrm{if $j=1$,}\\
0 & \quad\quad\quad\quad\quad\quad\quad\textrm{Otherwise.}\\
\end{array}\right.\nonumber
\end{eqnarray}
\item If $0<i<M$,
\setlength\arraycolsep{0.1em}
\medmuskip=-1mu
\thinmuskip=-1mu
\begin{eqnarray}\label{Pij_2_proof}
P_{i,j}=\left\{\begin{array}{ll}
1-\frac{{A_{i,i}}^m-{A_{i+1,i}}^m}
{\left\{1-\frac{\pi \left({R_1}+\left(i-1\right)v\right)^2}{S}\right\}^m-\left\{1-\frac{\pi \left(R_1+iv\right)^2}{S}\right\}^m} & \textrm{if $j=i-1$,}\\
\frac{{A_{i,i}}^m-{A_{i+1,i}}^m-{A_{i+1,i}}^m+{A_{i+1,i+1}}^m}
{\left\{1-\frac{\pi \left({R_1}+\left(i-1\right)v\right)^2}{S}\right\}^m-\left\{1-\frac{\pi \left(R_1+iv\right)^2}{S}\right\}^m} & \textrm{if $j=i$,}\\
\frac{{A_{i,i+1}}^m-{A_{i+1,i+1}}^m}
{\left\{1-\frac{\pi \left({R_1}+\left(i-1\right)v\right)^2}{S}\right\}^m-\left\{1-\frac{\pi \left(R_1+iv\right)^2}{S}\right\}^m} & \textrm{if $j=i+1$,}\\
0 & \textrm{Otherwise.}\\
\end{array}\right.\nonumber
\end{eqnarray}
\item If $i=M$,
\setlength\arraycolsep{0.1em}
\medmuskip=-1mu
\thinmuskip=-1mu
\begin{eqnarray}\label{Pij_3_proof}
P_{M,j}=\left\{\begin{array}{ll}
\frac{{A_{M,M-1}}^m-{A_{M,M}}^m}
{\left(1-\frac{\pi \left({R_1}+ M v \right)^2}{S}\right)^m} & \quad\quad\quad\quad\quad\quad\textrm{if $j=M-1$,}\\
\frac{{A_{M,M}}^m}
{\left(1-\frac{\pi \left({R_1}+ M v \right)^2}{S}\right)^m} & \quad\quad\quad\quad\quad\quad\textrm{if $j=M$,}\\
0 & \quad\quad\quad\quad\quad\quad\textrm{Otherwise.}\\
\end{array}\right.\nonumber
\end{eqnarray}
\end{itemize}

\subsection{Charging Probability $p_c$ of \eqref{p_c}}

Given that there are $i$ WCSs within $R_1$ from a node,
the probability that the node is charged by one of the WCSs $p_c(i)$ is
\begin{align}\label{p_c_derivation_1}
p_c(i)&=1-\left[1-\sum_{l=0}^{n-1}\min\left[1,\frac{u}{l+1}\right]f \left(l;n-1, \frac{\pi{R_1}^2}{S}\right)\right]^i\nonumber\\
&=1-\Gamma^i,
\end{align}
where $f(n;k,p)$ is the probability density function of the binomial distribution with parameters $n$, $k$ and $p$, and
\begin{eqnarray}\label{p_c_derivation_2}
\Gamma=1-F\left(u-2;n-1,\frac{\pi{R_1}^2}{S}\right)-\frac{u\left(1-F\left(u-1;n,\frac{\pi{R_1}^2}{S}\right)\right)}{n \pi R^2}.
\end{eqnarray}
The probability that there are $i$ WCSs within $R_1$ from a node is $f\left(i; m,\frac{\pi {R_1}^2}{S}\right)$.
Therefore, the charging probability $p_c$ is
\begin{eqnarray}\label{p_c_derivation_3}
p_c=\frac{\sum_{i=1}^m p_c(i)f\left(i;m,\frac{\pi {R_1}^2}{S}\right)}{1-\left(1-\frac{\pi {R_1}^2}{S}\right)^m}
\end{eqnarray}
The denominator of \eqref{p_c_derivation_3} represents the probability that the node is in one of the WCS's charging regions.
After substituting \eqref{p_c_derivation_1} into \eqref{p_c_derivation_3}, the charging probability $p_c$ becomes
{\setlength\arraycolsep{0.1em}\begin{align}\label{p_c_derivation_4}
p_c&=1-\frac{\sum_{i=1}^m  \binom{m}{i} \left(A\frac{\pi {R_1}^2}{S}\right)^i \left(1-\frac{\pi {R_1}^2}{S}\right)^{m-i}}{1-\left(1-\frac{\pi {R_1}^2}{S}\right)^m}\nonumber\\
&=\frac{1-\sum_{i=0}^m  \binom{m}{i} \left(A\frac{\pi {R_1}^2}{S}\right)^i \left(1-\frac{\pi {R_1}^2}{S}\right)^{m-i}}{1-\left(1-\frac{\pi {R_1}^2}{S}\right)^m}\nonumber\\
&=\frac{1-\left(A\frac{\pi {R_1}^2}{S}+1-\frac{\pi {R_1}^2}{S}\right)^m}{1-\left(1-\frac{\pi {R_1}^2}{S}\right)^m}.
\end{align}
After inserting \eqref{p_c_derivation_2} into \eqref{p_c_derivation_4}, we complete the proof.

\subsection{Transmission Probability $p_t$ of \eqref{p_t}}
Assume that a node is active.
The node consumes one unit of energy
when its mode is that of a transmitter, and there is at least one receiver within $r$:
{\setlength\arraycolsep{0.1em}
\begin{eqnarray}\label{p_t_derivation_1}
p_t&=&q\sum_{i=0}^{n-1} \left\{1-\left(1-\frac{\pi r^2}{S}\right)^i\right\}\binom{n-1}{i}
(1-q)^i q^{n-1-i}\nonumber\\
&=&q -q\sum_{i=0}^{n-1}\binom{n-1}{i}
\left\{(1-q)\left(1-\frac{\pi r^2}{S}\right)\right\}^i q^{n-1-i}
\nonumber\\
&=&q-q\left\{(1-q)\left(1-\frac{\pi r^2}{S}+q\right\}\right\}^{n-1}\nonumber\\
&=&q\left[1-\left\{1-(1-q)\frac{\pi r^2}{S}\right\}^{n-1} \right]\nonumber
\end{eqnarray}

\subsection{Proof of Proposition 1}

According to \cite{Platis98}, the CCDF of $T_I$ is
\begin{eqnarray}
\Pr\left[T_{I}>t\right]=\boldsymbol{p_0}\boldsymbol{P}^{t-1}\boldsymbol{1} \label{T_I_dist}
\end{eqnarray}
Assume that matrix $\boldsymbol{P}$ \eqref{P} is invertible\footnote{{It is a reasonable assumption that the transition probability, expressed as a row vector in $\boldsymbol{P}$,  is independent of the current location status $d$ of \eqref{partition} unless speed is infinite, 
$\boldsymbol{P}$ is likely to be a full rank matrix guaranteeing the existence of $M$ eigenvalues. 
It is also verified numerically under numerous combinations of parameter settings.}},
it can be diagonalized as follows:
\begin{align}\label{EVD}
\boldsymbol{P}&=\boldsymbol{V}\boldsymbol{D}\boldsymbol{V^{-1}}\nonumber\\
&=\left(\begin{array}{cccc}
\boldsymbol{a_1} & \boldsymbol{a_2} & \cdots & \boldsymbol{a_{M}}
\end{array}\right)
\left(\begin{array}{cccc}
\lambda_1 & 0 & \cdots & 0\\
0 & \lambda_2 & \cdots & 0\\
\vdots & \vdots & \ddots & \vdots\\
0 & 0 & \cdots & \lambda_{M}
\end{array}\right)
\left(\begin{array}{c}
\boldsymbol{b_1^T}\\
\boldsymbol{b_2^T} \\
\vdots \\
\boldsymbol{b_{M}^T}
\end{array}\right).
\end{align}
Therefore,  $\boldsymbol{P}^{t-1}$ is
\begin{eqnarray}\label{EVD_n}
\boldsymbol{P}^{t-1}
&=&\left(\begin{array}{cccc}
\boldsymbol{a_1} & \boldsymbol{a_2} & \dots & \boldsymbol{a_{M}}
\end{array}\right)
\left(\begin{array}{cccc}
{\lambda_1}^{t-1} & 0 & \dots & 0\\
0 & {\lambda_2}^{t-1} & \dots & 0\\
\vdots & \vdots & \ddots & \vdots\\
0 & 0 & \dots & {\lambda_{M}}^{t-1}
\end{array}\right)
\left(\begin{array}{c}
\boldsymbol{b_1^T}\\
\boldsymbol{b_2^T} \\
\vdots \\
\boldsymbol{b_{M}^T}
\end{array}\right)\nonumber\\
&=&\boldsymbol{a_1} \boldsymbol{b_1^T} {\lambda_1}^{t-1}+\boldsymbol{a_2} \boldsymbol{b_2^T} {\lambda_2}^{t-1}+\cdots+\boldsymbol{a_{M}} \boldsymbol{b_{M}^T} {\lambda_{M}}^{t-1}\nonumber\\
&=&\sum_{i=1}^{M}\boldsymbol{G_i}\cdot (\lambda_i)^{t-1},
\end{eqnarray}
where $\boldsymbol{G_i}=\boldsymbol{a_i} \boldsymbol{b_i^T}$ are $M \times M$ matrices of
which the sum is an identity matrix $\left(\sum_{i=1}^{M}\boldsymbol{G_i}=\boldsymbol{I}\right)$.
Using \eqref{EVD_n},
\eqref{T_I_dist} is rewritten~as 
\begin{eqnarray}\label{Proposition_1_1_proof}
\Pr\left[T_{I}>t\right]&=&\sum_{i=1}^{M}\boldsymbol{p_0}\boldsymbol{G_i}\boldsymbol{1} {\lambda_i}^{t-1}
=\sum_{i=1}^{M} \gamma_i \cdot {\lambda_i}^{t-1}.\nonumber
\end{eqnarray}
According to \cite{Meyer00}, every eigenvalue of an irreducible substochastic matrix is less than $1$. 
Matrix $\boldsymbol{P}$ \eqref{P} is a substochastic matrix 
because every row sum is $1$ except the first one due to a strictly positive transition probability $P_{1,0}$.
Consequently, $\lambda_{i}$ is smaller than one for every $i$.

\subsection{Proof of Proposition 2}

As the battery capacity $L$ goes infinity, this Markov process \eqref{Q} becomes batch Markovian arrival process (BMAP).
The stochastic process of BMAP can be described by the mean steady-state
arrival rate $\bar{\lambda}$.
According to \cite{Bolch06}, the authors explain how to derive $\bar{\lambda}$.
First, an infinite generator $\boldsymbol{D}$ is calculated as follows:
\begin{align}
&\boldsymbol{D}=\boldsymbol{B_0}+\boldsymbol{A_2}+\boldsymbol{A_3}+\cdots+\boldsymbol{A_{E-1}}\nonumber\\
&=
\left(\begin{array}{ccccc}
-P_{0,1}& P_{0,1}                  & 0                  & \cdots   & 0\\
P_{1,0}      &-P_{1, 0}-P_{1,2} & P_{1,2}              & \cdots   & 0\\
0                     & P_{2,1}                  &-P_{2,1}-P_{2,3} &  \cdots    & 0\\
\vdots                     & \vdots                 &\vdots  & \ddots &     \vdots\\
0                     & 0                       & 0                  & \cdots  & -P_{M, M-1}\\
\end{array}\right)
\end{align}
Let us denote by $\phi_{k}$ steady state probability that the relative distance $d$ is $k$.
We make the following row vector $\boldsymbol{\phi}$:
\setlength\arraycolsep{0.1em}\begin{align}
\boldsymbol{\phi}
&=\left[\begin{array}{ccccc}
\phi_{0}, & \phi_{1}, & \phi_{2}, & \cdots, & \phi_{M}\\
\end{array}\right]\nonumber\\
&=\left[\begin{array}{ccccc}
\sum_{j=0}^{L} \pi_{0,j}, & \sum_{j=0}^{L} \pi_{1,j}, & \sum_{j=0}^{L} \pi_{2,j}, & \cdots, & \sum_{j=0}^{L} \pi_{M,j}\\
\end{array}\right]\nonumber
\end{align}
It can be obtained by solving the following equations:
{\setlength\arraycolsep{0.1em}\begin{eqnarray}\label{Equation_1}
\boldsymbol{\phi}\boldsymbol{D}=0, \quad
\boldsymbol{\phi}\boldsymbol{1}=1.
\end{eqnarray}
We get $\phi_{k}$ as follows:
\setlength\arraycolsep{0.1em}
\medmuskip=-2mu
\thinmuskip=-2mu
\begin{align}\label{phi}
\phi_{k}=
\left\{\begin{array}{ll}
1-\left(1-\frac{\pi {R_1}^2}{S}\right)^m & \textrm{if $k=0$}\\
\left(1-\frac{\pi \left({R_1}+ M v \right)^2}{S}\right)^m & \textrm{if $k=M$}\\
\left\{1-\frac{\pi \left({R_1}+\left(k-1\right)v\right)^2}{S}\right\}^m-\left\{1-\frac{\pi \left(R_1+kv\right)^2}{S}\right\}^m & \textrm{Otherwise}\\
\end{array}\right.
\end{align}
From \eqref{phi}, we derive the mean steady-state
arrival rate $\bar{\lambda}$:
\setlength\arraycolsep{0.1em}
\medmuskip=-1mu
\thinmuskip=-1mu
\begin{eqnarray}\label{mean_arrival_rate}
\bar{\lambda}=\boldsymbol{\phi}\left(\sum_{k=1}^{E}k \boldsymbol{A_{k+1}}\right)\boldsymbol{1}
=p_c\left\{1-\left(1-\frac{\pi {R_1}^2}{S}\right)^m\right\}\left(\sum_{k=1}^{E}k\beta(k)\right)\nonumber
\end{eqnarray}
If $\bar{\lambda}<p_t$, the active probability $P_{\mathrm{on}}$ is
\begin{eqnarray}\label{active_Probability_infinite_battery}
P_{\mathrm{on}}=\frac{\bar{\lambda}}{p_t}=\frac{p_c}{p_t}\left\{1-\left(1-\frac{\pi {R_1}^2}{S}\right)^m\right\}\left(\sum_{k=1}^{E}k\beta(k)\right).
\end{eqnarray}
Otherwise, $P_{\mathrm{on}}=1$. After inserting \eqref{active_Probability_infinite_battery} into \eqref{Lemma1}, we obtain Proposition 2 \eqref{Proposition_2}.

\subsection{Proof of Proposition 3}

For the first step, we check the ratio $\frac{p_c}{p_t}$ as the number $n$ increases:
\begin{align}\label{Proposition3_proof1}
\frac{p_c}{p_t}&\approx
\frac{1-\left(1-\frac{u}{n}\right)^m}{q\left(1-(1-\frac{\pi R_1^2}{S})^m\right)\left(1-e^{-\frac{\pi}{4}\left(1-q\right)}\right)}\nonumber\\
&=\left\{\begin{array}{ll}
\frac{m}{n}\frac{u}{q\left(1-(1-\frac{\pi R_1^2}{S})^m\right)\left(1-e^{-\frac{\pi}{4}\left(1-q\right)}\right)} & \textrm{if $m=O\left(n\right)$}\\
\frac{1}{q\left(1-(1-\frac{\pi R_1^2}{S})^m\right)\left(1-e^{-\frac{\pi}{4}\left(1-q\right)}\right)} & \textrm{Otherwise.}\\
\end{array}\right.
\end{align}
We already proved the upper bound according to Proposition 2 \eqref{Proposition_2} as follows:
\begin{eqnarray}\label{Upper_bound_Proposition3}
\Lambda\leq \Lambda_{\mathrm{upper}}
&=&\Theta\left(\min\left(1,\frac{m}{n}\right){c_1}^{\min\left(1,\frac{m}{n}\right)}\right)
\end{eqnarray}
where $c_1=e^{-\frac{\pi \cdot u}{4\cdot a}\left(\sum_{k=1}^{E}k\beta(k)\right)}$.

In order to derive the lower bound, consider each WCS only delivers one unit of energy to a node in a time slot ($E=1$).
Since the submatrices $\boldsymbol{A_{2}}$ and $\boldsymbol{A_{3}}$ in the generating matrix $\boldsymbol{Q}$ \eqref{Q} is null matrices,
the Markov process becomes finite Quasi-Birth-Death Process (QBD).
In \cite{Akar1997}, the authors showed that the steady state probability vector $\boldsymbol{\pi_k}$ \eqref{each_partition_steady_state_probability}
of finite QBD can be expressed in a matrix geometric form:
\begin{eqnarray}\label{Steady_state_Matrix_Geometric}
\boldsymbol{\pi_k}= {\boldsymbol{v_1}}{\boldsymbol{R_1}}^k+{\boldsymbol{v_2}}{\boldsymbol{R_2}}^{L-k}.
\end{eqnarray}
Here, matrices $\boldsymbol{R_1}$ and $\boldsymbol{R_2}$ are
\begin{eqnarray}
&&\boldsymbol{R_1}=-\boldsymbol{A_2}\left(\boldsymbol{A_1}+\eta \boldsymbol{A_0}\right)^{-1}\label{R1}\\
&&\boldsymbol{R_2}=-\boldsymbol{A_0}\left(\boldsymbol{A_1}+\boldsymbol{A_0}\boldsymbol{G}\right)^{-1}\label{R2}
\end{eqnarray}
where $\eta$ is the spectral radius of $\boldsymbol{R_1}$,
and $\boldsymbol{G}$ is the square matrix that the every element of the first column is one and the others are zero.
Detailed derivations of $R_1$ and $R_2$ are in \cite{Ramswami1986}.
Row vectors $\boldsymbol{v_1}$ and $\boldsymbol{v_2}$ satisfy the following conditions:
\setlength\arraycolsep{0.1em}\begin{align}
\medmuskip=-1mu
\thinmuskip=-1mu
&\left[\begin{array}{cc}
\boldsymbol{v_1} &
\boldsymbol{v_2}\\
\end{array}\right]
\left[\begin{array}{cc}
\boldsymbol{B_1}+\boldsymbol{R_1}\boldsymbol{A_0} &
{\boldsymbol{R_1}}^{L-1}\left(\boldsymbol{A_0}+\boldsymbol{R_1}\left(\boldsymbol{A_0}+\boldsymbol{B_0}\right)\right) \\
{\boldsymbol{R_2}}^{L-1}\left(\boldsymbol{R_2}\boldsymbol{B_0}+\boldsymbol{A_0}\right) &
\boldsymbol{A_0}+\boldsymbol{A_1}+\boldsymbol{R_2}\boldsymbol{A_0}\\
\end{array}\right]
=\boldsymbol{0}\label{Boundary_condition}\\
&\left(\boldsymbol{v_1}\sum_{i=0}^L {\boldsymbol{R_1}}^i+\boldsymbol{v_1}\sum_{i=0}^L {\boldsymbol{R_1}}^i\right)\boldsymbol{1}=1\label{normarization_condition}
\end{align}
The boundary condition \eqref{Boundary_condition} is derived by inserting \eqref{Steady_state_Matrix_Geometric} into
the first and last columns of the balance equation \eqref{Equation_steady_state_probability},
and condition \eqref{normarization_condition} means the summation of the entire steady state probabilities is one.

\begin{figure*} [t]
\setlength\arraycolsep{0.1em}
\medmuskip=-1mu
\thinmuskip=-1mu
\begin{eqnarray}\small\label{X1X2X3}
X_1&=&\left|\begin{array}{cc}
\frac{p_t}{p_c}\frac{P_{0,1}+p_c}{P_{1,0}}+1    & \frac{p_t}{p_c}\frac{P_{0,1}+p_c}{P_{1,0}} \frac{P_{1,2}}{P_{2,1}}\\
\frac{p_t}{p_c}\frac{P_{0,1}+p_c}{P_{1,0}}+1    & \frac{p_t}{p_c}\frac{P_{0,1}+p_c}{P_{1,0}} \frac{P_{1,2}}{P_{2,1}}+\frac{p_t}{P_{2,1}}+1\\
\end{array}\right|,\quad
X_2=\left|\begin{array}{cc}
\frac{p_t}{p_c}\frac{P_{0,1}}{P_{1,0}}    & \frac{p_t}{p_c}\frac{P_{0,1}}{P_{1,0}} \frac{P_{1,2}}{P_{2,1}}\\
\frac{p_t}{p_c}\frac{P_{0,1}+p_c}{P_{1,0}}+1    & \frac{p_t}{p_c}\frac{P_{0,1}+p_c}{P_{1,0}} \frac{P_{1,2}}{P_{2,1}}+\frac{p_c}{P_{2,1}}+1\nonumber\\
\end{array}\right|,\quad\\
X_3&=&\left|\begin{array}{cc}
\frac{p_t}{p_c}\frac{P_{0,1}}{P_{1,0}}    & \frac{p_t}{p_c}\frac{P_{0,1}}{P_{1,0}} \frac{P_{1,2}}{P_{2,1}}\\
\frac{p_t}{p_c}\frac{P_{0,1}+p_c}{P_{1,0}}+1    & \frac{p_t}{p_c}\frac{P_{0,1}+p_c}{P_{1,0}} \frac{P_{1,2}}{P_{2,1}}\\
\end{array}\right|,
\end{eqnarray}
where $\left|\begin{array}{cc}
a & b\\
c & d\\
\end{array}\right|=ad-bc$.
\hrulefill
\end{figure*}

From the equation \eqref{Steady_state_Matrix_Geometric}, the active probability $P_{\mathrm{on}}$  \eqref{Active_Probability} is rewritten as follows:
\setlength\arraycolsep{0.1em}\begin{eqnarray}
P_{\mathrm{on}}=1-(\boldsymbol{v_1}+\boldsymbol{v_2}{\boldsymbol{R_2}}^L)\boldsymbol{1}\label{Active_Probability_matrix_geometric}
\end{eqnarray}
Since the active probability $P_{\mathrm{on}}$ is a non-decreasing function of the battery capacity $L$,
we make the following inequality condition:
\setlength\arraycolsep{0.1em}\begin{eqnarray}
P_{\mathrm{on}}&\geq& 1-(\boldsymbol{v_1}+\boldsymbol{v_2}{\boldsymbol{R_2}})\boldsymbol{1}\nonumber\\
&=&\left(\boldsymbol{v_1}\sum_{i=0}^1 {\boldsymbol{R_1}}^i+\boldsymbol{v_1}\sum_{i=0}^1 {\boldsymbol{R_1}}^i\right)\boldsymbol{1}-(\boldsymbol{v_1}+\boldsymbol{v_2}{\boldsymbol{R_2}})\boldsymbol{1}\nonumber\\
&=&\left(\boldsymbol{v_1}\boldsymbol{R_1}+\boldsymbol{v_2}\right)\boldsymbol{1}\label{Active_Probability_inequality_1}
\end{eqnarray}
All elements in matrix $\boldsymbol{R_1}$ is zero except the first row, and the first element of $\boldsymbol{v_1}$ is zero.
Therefore,  $\boldsymbol{v_1}\boldsymbol{R_1}$ becomes zero and the above inequality \eqref{Active_Probability_inequality_1} becomes
\setlength\arraycolsep{0.1em}\begin{eqnarray}
P_{\mathrm{on}}&\geq&\boldsymbol{v_2}\boldsymbol{1}=\sum_{i=0}^M v_{2,i}\geq v_{2,0}\label{Active_Probability_inequality_2}
\end{eqnarray}
From the condition \eqref{Boundary_condition}, we make the following relations between $\boldsymbol{v_1}$ and $\boldsymbol{v_2}$,
\setlength\arraycolsep{0.1em}\begin{eqnarray}
&&\boldsymbol{v_1}=-\boldsymbol{v_2}\left(\boldsymbol{R_2} + p_t \boldsymbol{B_0}^{-1}\right),\label{relation1}\\
&&\boldsymbol{v_1}+\boldsymbol{v_2}\left(\boldsymbol{I}+\boldsymbol{R_2}\right)=\boldsymbol{\phi}.\label{relation2}
\end{eqnarray}
After inserting \eqref{relation1} into \eqref{relation2}, we derive $\boldsymbol{v_2}$ as 
\setlength\arraycolsep{0.1em}\begin{eqnarray}
\boldsymbol{v_2}=\boldsymbol{\phi}\left(\boldsymbol{I} - p_t \boldsymbol{B_0}^{-1}\right)^{-1}.\label{v2}
\end{eqnarray}
According to numerical verifications, it is checked that $v_{2,0}$ is an increasing function of the resolution factor $M$.
When $M=3$, the vector $\boldsymbol{v_2}$ becomes:
\setlength\arraycolsep{0.1em}\begin{align}
v_{2,0}
&=\boldsymbol{\phi} \frac{1}{(\frac{p_t}{p_c}+1)X_1-\frac{p_t}{p_c}X_2+\frac{p_t}{p_c}X_3}\left(\begin{array}{c}
X_1           \\
-X_2          \\
X_3           \\
\end{array}\right)\nonumber\\
&=\frac{
\phi_1 X_1-\phi_2 X_2+\phi_3 X_3}{(\frac{p_t}{p_c}+1)X_1-\frac{p_t}{p_c}X_2+\frac{p_t}{p_c}X_3}\nonumber\\
&>\frac{p_c}{p_t}\frac{\left(\phi_1-\phi_2\frac{X_2}{X_1}+\phi_3\frac{X_3}{X_1}\right)}{2},
\label{v2_M}
\end{align}
where $X_1$, $X_2$ and $X_3$ are described in \eqref{X1X2X3}.
As $n$ increases, the ratios  $\frac{X_2}{X_1}$ and $\frac{X_2}{X_1}$ reduce to zero.
From the inequalities \eqref{Active_Probability_inequality_2} and \eqref{v2_M}, the active probability $P_{\mathrm{on}}$ is
\begin{eqnarray} \label{Active_Probability_inequality_4}
P_{\mathrm{on}}&>&\frac{p_c}{p_t}\frac{\phi_1}{2}=\frac{p_c}{p_t}\frac{\left\{1-\left(1-\frac{\pi {R_1}^2}{S}\right)^m\right\}}{2}.
\end{eqnarray}
We can derive the lower bound of the throughput $\Lambda$ as
\begin{eqnarray}\label{lower_bound_Proposition3}
\Lambda>\Lambda_{\textrm{lower}}=\Theta\left(\min\left(1,\frac{m}{n}\right){c_2}^{\min\left(1,\frac{m}{n}\right)}\right),
\end{eqnarray}
where $c_2=e^{-\frac{\pi \cdot u}{8\cdot a}}$.
From the upper bound \eqref{Upper_bound_Proposition3} and lower bound \eqref{lower_bound_Proposition3}, we complete to prove Proposition 3.

\bibliographystyle{IEEE}

\end{document}